\documentclass[manuscript]{emulateapj}
\usepackage{float}
\usepackage{epsfig}     
\usepackage{amsmath}
\usepackage{color}
\usepackage{courier}
\usepackage{comment}
\usepackage{subfigure}
\usepackage{enumitem}
\usepackage{soul}

\slugcomment{Accepted for publication in the Astrophysical Journal Letters (ApJL)}
\shorttitle{Cosmic Star-Formation in Galaxy Protoclusters}
\shortauthors{Y.-K. Chiang et al.}

\begin{document}

\title{Galaxy Protoclusters as Drivers of Cosmic Star-Formation History in the First 2 Gyr}
\author{Yi-Kuan Chiang\altaffilmark{1}, Roderik A. Overzier\altaffilmark{2}, Karl Gebhardt\altaffilmark{3}, Bruno Henriques\altaffilmark{4}}

\altaffiltext{1}{Department of Physics \& Astronomy, Johns Hopkins University, 3400 N. Charles Street, Baltimore, MD 21218, USA, ykchiang@jhu.edu}
\altaffiltext{2}{Observat\'orio Nacional, Rua Jos\'e Cristino, 77. CEP 20921-400, S\~ao Crist\'ov\~ao, Rio de Janeiro-RJ, Brazil}
\altaffiltext{3}{Department of Astronomy, University of Texas at Austin, 1 University Station C1400, Austin, TX 78712, USA}
\altaffiltext{4}{Department of Physics, Institute for Astronomy, ETH Zurich, CH-8093 Zurich, Switzerland}

\begin{abstract}
Present-day clusters are massive halos containing mostly quiescent galaxies, while distant protoclusters are extended structures containing numerous star-forming galaxies. We investigate the implications of this fundamental change in a cosmological context using a set of $N$-body simulations and semi-analytic models. We find that the fraction of the cosmic volume occupied by all (proto)clusters increases by nearly three orders of magnitude from $z=0$ to $z=7$. We show that (proto)cluster galaxies are an important, and even dominant population at high redshift, as their expected contribution to the cosmic star-formation rate density rises (from $1\%$ at $z=0$) to $20\%$ at $z=2$ and $50\%$ at $z=10$. Protoclusters thus provide a significant fraction of the cosmic ionizing photons, and may have been crucial in driving the timing and topology of cosmic reionization. Internally, the average history of cluster formation can be described by three distinct phases: at $z\sim10$--$5$, galaxy growth in protoclusters proceeded in an inside-out manner, with centrally dominant halos that are among the most active regions in the Universe; at $z\sim5$--$1.5$, rapid star formation occurred within the entire $10$--$20$ Mpc structures, forming most of their present-day stellar mass; at $z\lesssim1.5$, violent gravitational collapse drove these stellar contents into single cluster halos, largely erasing the details of cluster galaxy formation due to relaxation and virialization. Our results motivate observations of distant protoclusters in order to understand the rapid, extended stellar growth during Cosmic Noon, and their connection to reionization during Cosmic Dawn.
\end{abstract}


\section{Introduction}

Galaxy clusters today make a negligible contribution to the cosmic star-formation rate density (CSFRD) due to their low number density \citep{1993ApJ...407L..49B}, compact sizes \citep{1995ApJ...438..527G}, and quenched galaxy populations \citep{2004MNRAS.353..713K}. However, the same is not true for the cosmic stellar mass budget, as clusters are host to a large number of massive galaxies \citep{2011MNRAS.412..246V,2013A&A...557A..15V}.  Massive cluster galaxies, mostly ellipticals and S0s, are believed to have a  star-formation history (SFH) shifted to an earlier epoch compared to field galaxies \citep{2010MNRAS.404.1775T}, implying that the contribution from cluster progenitors to the CSFRD grows with redshift. Furthermore, at $z>1$ (proto)clusters are not yet confined to their present-day virial radius but spread out over the large-scale structures from which they form \citep{2013ApJ...779..127C}. The protocluster ``filling factor'', defined as the fraction of cosmic volume occupied by (proto)clusters, was thus significantly larger than it is today.

The physical processes that drive the rise and fall of star formation in cluster galaxies are not precisely known yet. Part of the problem is that once the final cluster has been assembled, some signatures of its formation history are erased due to merging, relaxation and virialization. Therefore, in recent years a great amount of effort has been spent to find and study high redshift protoclusters in which these processes can be directly observed \citep[][]{2007A&A...461..823V,2013ApJ...769...79W,2014ApJ...782L...3C,2015A&A...582A..30P,2016ApJ...826..114T}.

\begin{figure*}[t!]
     \begin{center}
        \subfigure{%
           \includegraphics[width=\textwidth]{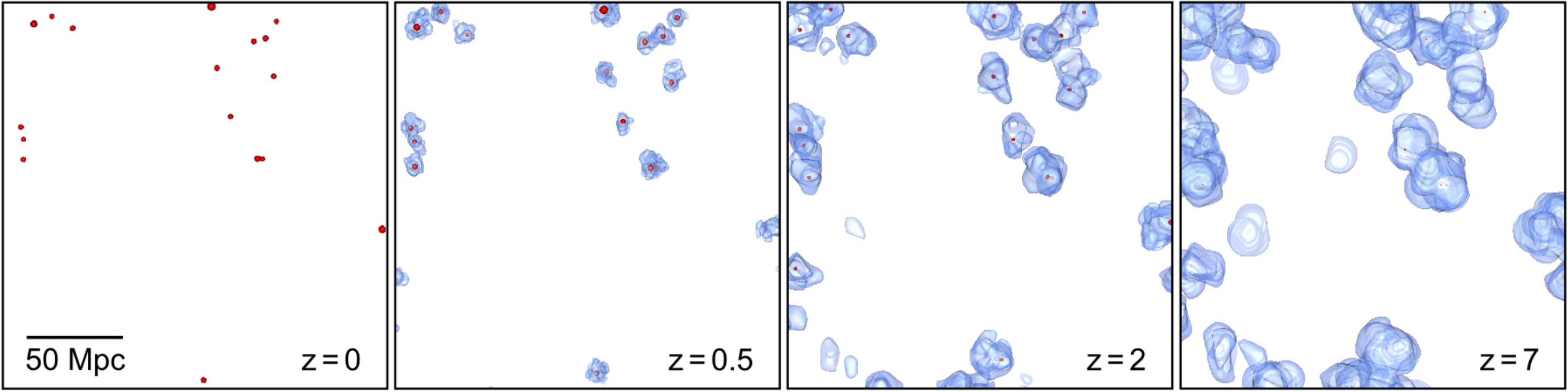}
        }\\ 
    \caption{\label{fig:popcorn}
    Cosmic volumes occupied by (proto)clusters of $M_{z=0}>10^{14}\ \rm M_{\odot}$ at $z=0$, 0.5, 2, and 7 in a slab of $200\times200\times50$ comoving Mpc$^3$. Blue surfaces indicate the Lagrangian boundaries of (proto)clusters. Red spheres indicate the virial radii of the most massive core halos.
     }%
    \end{center}
\end{figure*}

The CSFRD is one of the most important quantities in extragalactic astronomy, as it concisely summarizes the emergence and evolution of galaxies, and the radiation and heavy elements that they produce \citep[][]{2014ARA&A..52..415M}. To understand the role of galaxy clusters in the cosmic star-formation history, here we investigate the contribution to the CSFRD from galaxies in the ``field'', in protoclusters, and in the cores of protoclusters using two semi-analytic models. The main result is that the contribution from (proto)clusters to the CSFRD is expected to rise significantly from $1$\% at $z=0$ to $20\%$ at $z=2$ and $50\%$ at $z=10$. The structure of this Letter is as follows. We introduce the simulations in Section 2. We analyze the volume filling factors of protoclusters in Section 3. The contributions to the CSFRD and the internal evolution of (proto)clusters are presented in Sections 4 and 5, respectively. We discuss our results in Section 6.

\section{Simulations and Sample Construction}
\label{Simulations and Sample Construction}

To predict the cosmic volume and star formation associated with (proto)clusters\footnote{We use ``(proto)clusters'' as a shorthand for ``clusters and protoclusters'' in context that involves the epoch of $0 < z\lesssim 2$ when these two coexist.}, we use two recent semi-analytic models (SAMs) that have similar physics, but with their free parameters constrained in different ways. Our preferred SAM is that of \cite{2015MNRAS.451.2663H} (hereafter H15), whose free parameters were constrained using the observed galaxy stellar mass functions and mass-dependent quiescent fractions at $z=0$-$3$ (see the appendix in H15 for details). In the model, environmental effects naturally arise from the hierarchical assembly of halos and galaxies and from specific recipes that operate only in dense environments. H15 reproduces the clustering of local galaxies as function of stellar mass ($M_{\star}$) and color, as well as masses, star-formation rates (SFR)  and quiescent fractions at least up to $z=1$ \citep{2016arXiv161102286H}.  We also use the SAM of \cite{2013MNRAS.428.1351G} (hereafter G13), which was constrained only using the local galaxy stellar mass function. 

The two SAMs were run on the Millennium cosmological $N$-body simulation \citep{2005Natur.435..629S} scaled to the \textsc{Planck} (H15) and WMAP7 (G13) cosmologies following the techniqe in \cite{2010MNRAS.405..143A}. This yields an effective volumes of (480.3 Mpc $h^{-1})^3$ for H15 and (521.6 Mpc $h^{-1})^3$ for G13. The different cosmologies mainly impact our predictions for the (proto)cluster CSFRD and volume filling factor through changes in the number densities of objects: there are $40\%$ more $>10^{14}\ \rm M_{\odot}$ halos at $z=0$ in the \textsc{Planck} compared to the WMAP7 cosmology. To facilitate the comparison between the two models, in Section 3 and 4, we have therefore scaled our results to the \textsc{Planck} cluster abundance. Given the mass resolution of the simulations, the SFR and $M_{\star}$ presented in this work are based on galaxies with $M_{\star}>10^{8.5}\ \rm M_{\odot}$ ($1.5$--$2.5$ dex below the characteristic $M_{*}$). This corresponds to a limiting UV absolute magnitude of $-13$ ($-18.5$) at $z=0$ ($z=10$). A Salpeter initial mass function \citep{1955ApJ...121..161S} is assumed.

We select a mass-complete cluster sample of $M_{200}>10^{14}\ \rm M_{\odot}$ at $z=0$, leading to 3819 (2981) clusters for H15 (G13). $M_{200}$ is the mass enclosed by the (virial) radius $R_{200}$ within which the overdensity is 200 times the critical density. Throughout the paper, we consider $R_{200}$ to be an estimate of the boundary of the halo. Our cluster sample has a mean mass of $2\times 10^{14}\ \rm M_{\odot}$ at $z=0$, with a high tail extending to $2\times 10^{15}\ \rm M_{\odot}$.

A protocluster is defined as the collection of all the dark matter and baryons that will assemble into a $z=0$ cluster. The outermost boundary of these moving particles defines a three-dimensional shape that contains the Lagrangian volume of the system. For each $z=0$ cluster, we trace the merger trees of all the sub-halos within $R_{200}$ to high redshift, and refer to all the galaxies associated with these merger trees as members of the protocluster. We define a protocluster ``core'' as the most massive halo in a protocluster at any given epoch. This is motivated by the fact that some $z\sim2$ protoclusters already contain dominant central halos capable of hosting extended X-ray emission, Sunyaev-Zel'dovich effect \citep{1980ARA&A..18..537S} signal or massive quiescent galaxies.

\section{Volume Filling Factor of Protoclusters}

To highlight the role of protoclusters in cosmic structure formation, we first quantify the volumes that they occupy. In Fig.~\ref{fig:popcorn} we visualize a slab cutout ($200\times 200 \times 50$ Mpc$^3$) in several H15 snapshots from $z=0$--$7$. The snapshot at $z=0$ shows the locations of all the $M_{200}>10^{14}\ \rm M_{\odot}$ clusters (red spheres scaled to $R_{200}$). In the other panels, protocluster cores are indicated in the same manner, but now we also show the Lagrangian volumes occupied by the entire protocluster (blue surfaces). The protocluster volumes were defined such that matter which lies within the surface will become part of the cluster by $z=0$ while matter outside the surface will not\footnote{Evaluated by comparing two smoothed, normalized probability density fields in three-dimensional space, one for protocluster galaxies, one for non-protocluster galaxies. These two scalar fields are summed up with one of the signs flipped. The isosurfaces at the value of zero are then taken as their boundaries.}. Although protoclusters collapse with time, the cores grow as a result of mass assembly and virialization.

\begin{figure}[t]
     \begin{center}
        \subfigure{%
           \includegraphics[width=\columnwidth]{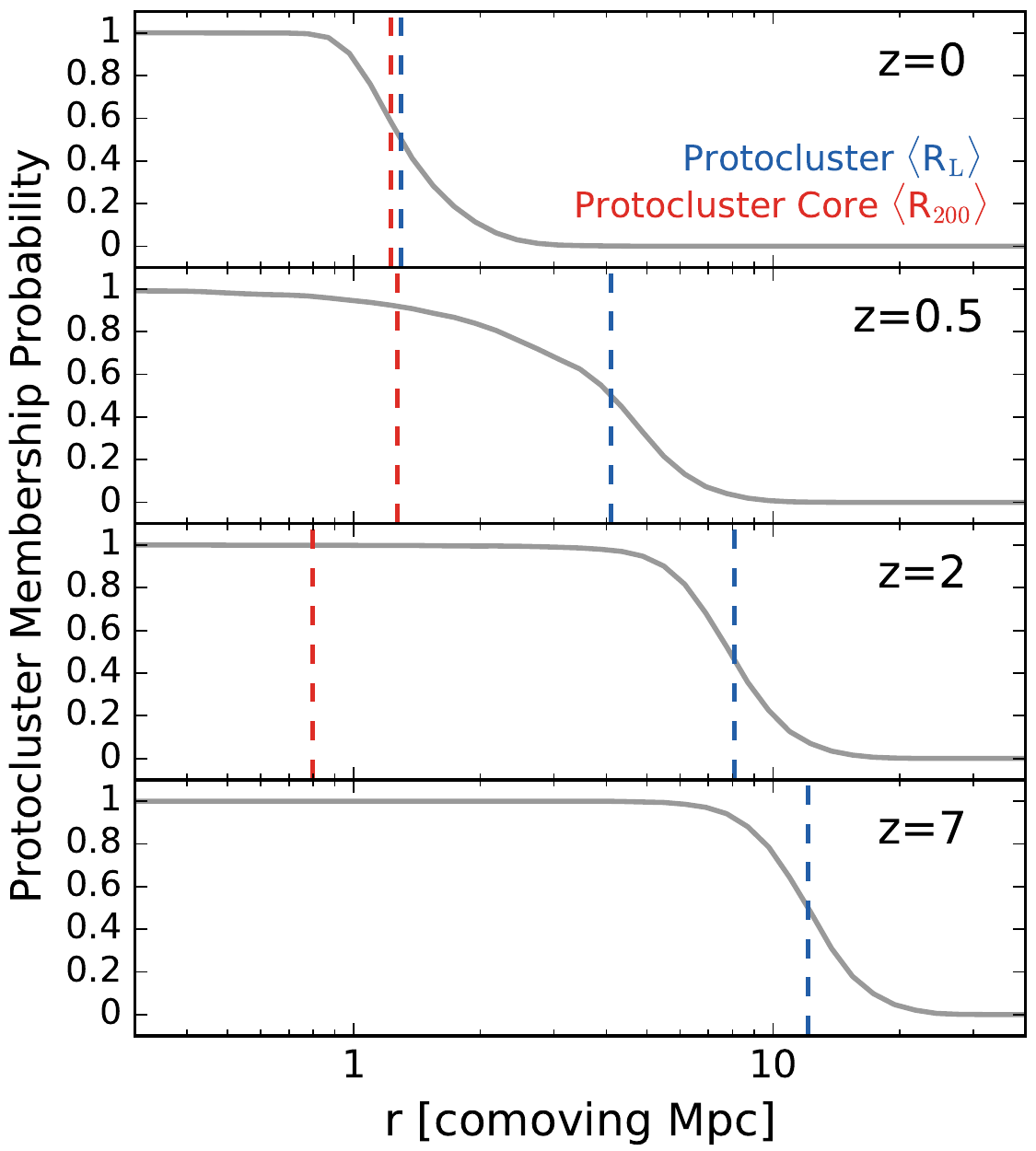}
        }\\ 
    \caption{\label{fig:density_profile}
    Stacked protocluster membership probability profiles. Solid curves show the probability for a test galaxy found at a given cluster-centric distance to become part of the stacked $z=0$ cluster. The mean Lagrangian radius $\langle R_{\rm L} \rangle$, and the mean $R_{200}$ of the protocluster cores are indicated with blue and red vertical dashed lines, respectively.}%
    \end{center}
\end{figure}

To simplify the volume calculations, we define a mean spherical Lagrangian radius $\langle R_{\rm L}\rangle$ as the boundary of a protocluster by taking the distance at which the membership probability drops to $50\%$. This is demonstrated in Fig.~\ref{fig:density_profile}, which shows the protocluster membership probability profiles as function of the distance to the stellar barycenter averaged over all the clusters in our sample. We note that $\langle R_{\rm L}\rangle$ is a total radius, not a half-mass radius. The smoothing of these profiles can be largely attributed to the spread in cluster mass within the sample, as opposed to a large departure from spherical symmetry. Both at $z=0$ and at $z\geq2$, these profiles fall off rather sharply as $\langle R_{\rm L}\rangle$ separates galaxies that are protocluster members from non-members. However, for an extended epoch of $z\sim0$--$1$, the boundary marked by $R_{\rm L}$ is less sharp (second panel from the top). This epoch corresponds to the post-turnaround, free-falling phase for the outermost shells of cluster material, and the details of this phase vary widely from cluster to cluster.

\begin{figure}[t]
     \begin{center}
        \subfigure{%
           \includegraphics[width=\columnwidth]{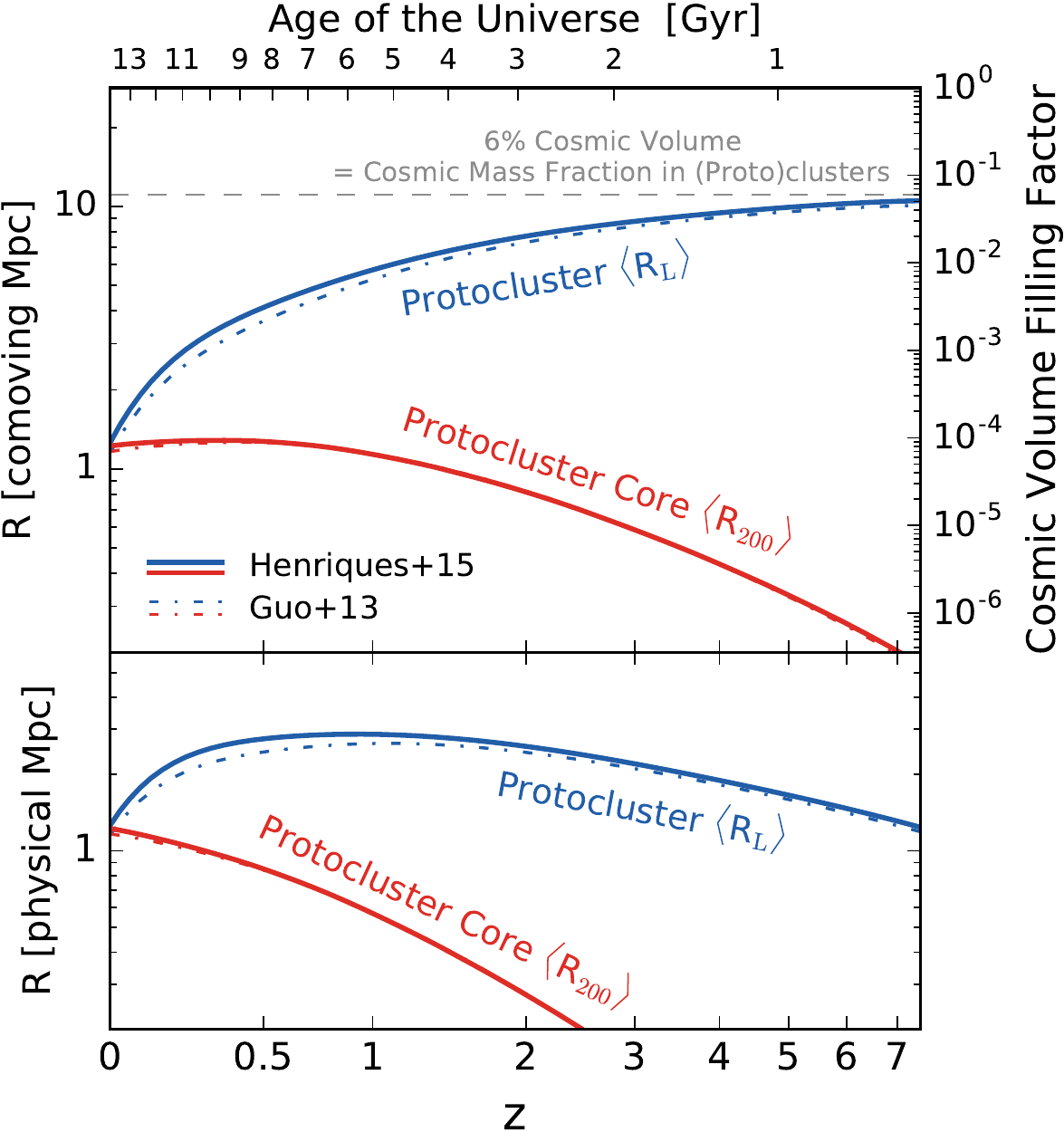}
        }\\ 
\caption{\label{fig:r_l}Average size evolution of protoclusters ($\langle R_{\rm L} \rangle$; blue curves), and protocluster cores ($\langle R_{\rm 200} \rangle$; red curves) in comoving (upper panel) and physical coordinates (lower panel). The corresponding cosmic volume filling factors are shown on the right-hand  $y$-axis in the upper panel.}
    \end{center}
\end{figure}

\begin{figure*}[t!]
     \begin{center}
     \label{comparison}
        \subfigure{%
           \includegraphics[width=0.725\textwidth]{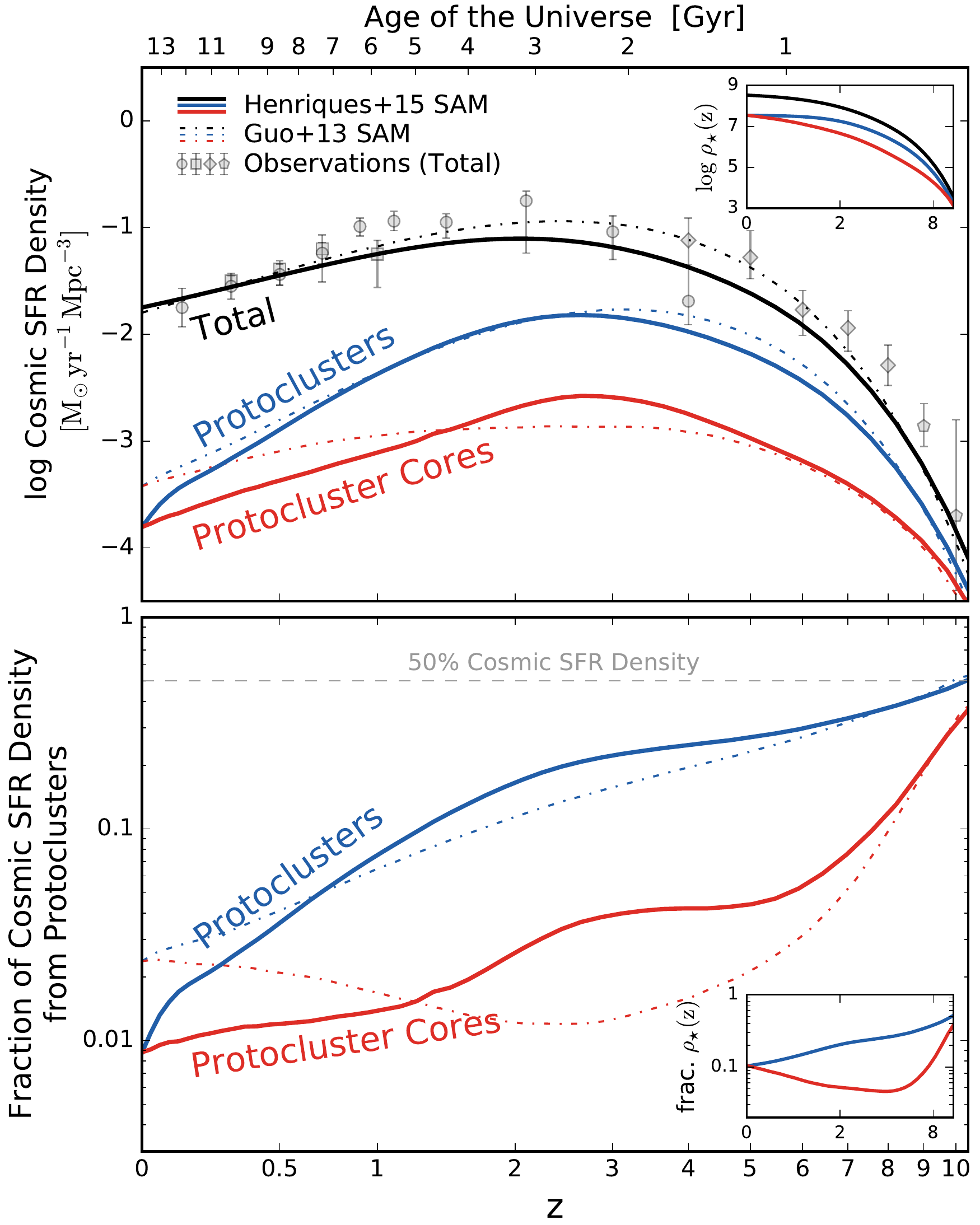}
        }
    \end{center}
    \caption{\label{fig:csfr}
    \textbf{Upper panel:} SFR density for all galaxies (black), protoclusters (blue), and cores (red). The associated stellar mass densities are shown in the inset. Data points show a set of observationally derived cosmic SFR density for comparison (circles: \cite{2012A&A...539A..31C}; squares: \cite{2005ApJ...619L..47S}; diamonds: \cite{2015ApJ...810...71F}, pentagons: \cite{2013ApJ...773...75O}). \textbf{Lower panel:} Fractional contributions to the total cosmic SFR density of protoclusters (blue) and protocluster cores (red). The associated stellar mass density fractions are shown in the inset.}%
\end{figure*}

Fig.~\ref{fig:r_l} shows the redshift evolution of $\langle R_{\rm L}  \rangle$. In comoving coordinates (upper panel), $\langle R_{\rm L} \rangle$ increases from 1.2 Mpc at $z=0$ to over 10 Mpc at $z=7$. In physical coordinates (lower panel), $\langle R_{\rm L} \rangle$ reaches a maximum of $3$ Mpc at $z\sim1$ after which it decreases as the cluster material turns around under self-gravity. The increasing comoving sizes of (proto)clusters with redshift naturally lead to an increasing cosmic volume filling factor (right-hand vertical axis in Fig.~\ref{fig:r_l}). Today, collapsed clusters occupy a negligible fraction ($\sim 10^{-4}$) of cosmic volume. However, at $z=7$ this fraction has increased by a factor of $500$ to $5\%$. An almost exact asymptotic boundary condition can be calculated for $z\rightarrow \infty$ when the density of the Universe was nearly uniform and the volume filling factor was simply the cosmic mass fraction in (proto)clusters ($6\%$, independent of redshift).

Fig.~\ref{fig:r_l} also shows that the cosmic volume enclosed within the mean virial radius of the protocluster core (red lines) quickly becomes insignificant toward higher redshift. This highlights the importance of studying the entire protocluster and not just the most massive, central halo in the cluster merger tree.

\begin{figure*}[t!]
     \begin{center}
     \label{comparison}
        \subfigure{%
           \includegraphics[width=0.78\textwidth]{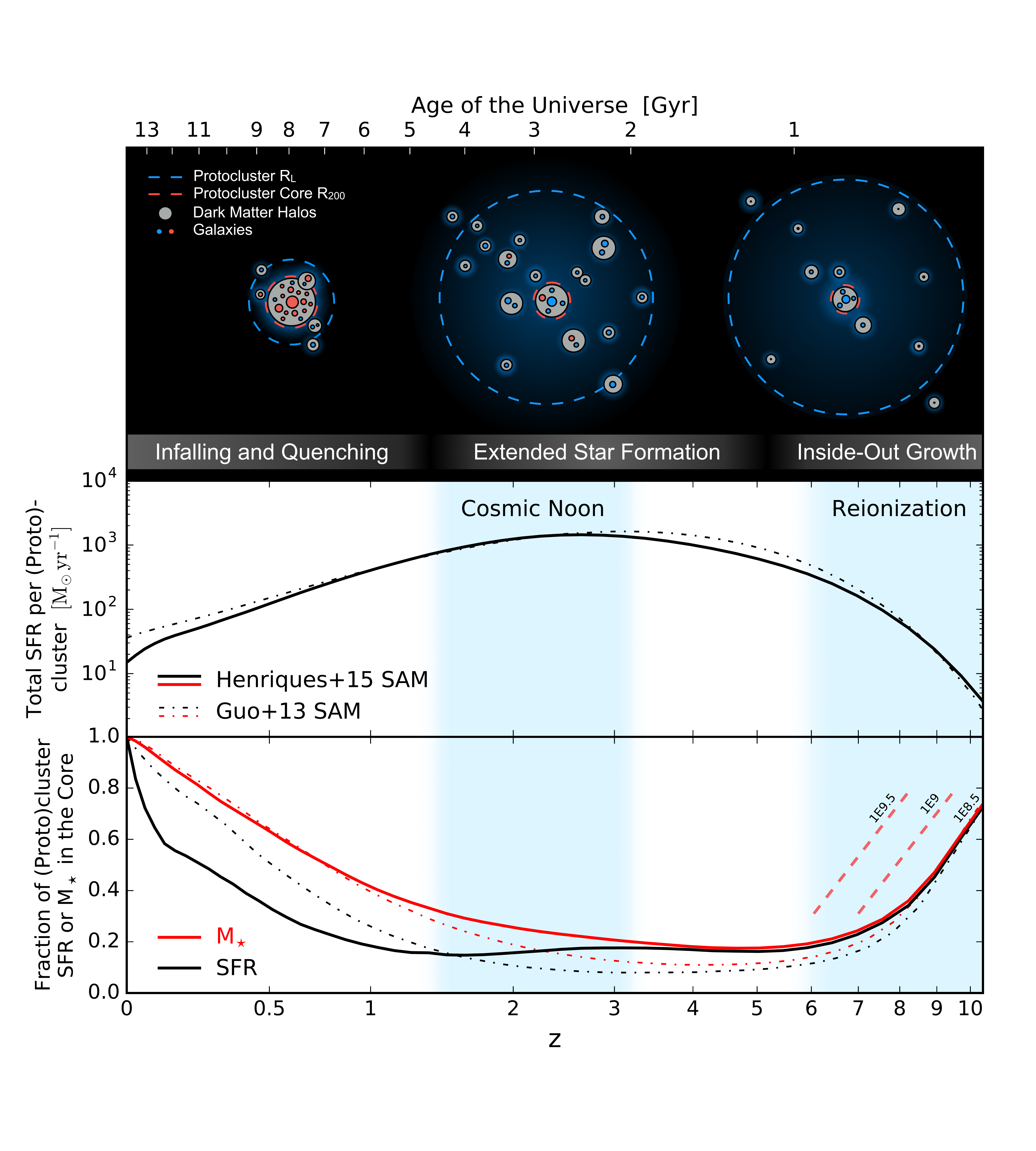}
        }
    \end{center}
    \caption{\label{fig:internal_evolution}
    Average total SFR per protocluster (middle panel) and the fractions of the total SFR (black curves) and stellar mass (red curves) occurring in the core halo (lower panel). The dashed lines at high redshift (lower panel) illustrate the dependence of the limiting galaxy stellar mass on the upturn of core dominance. The rise and fall in the total SFR of protoclusters and the reversed trend found for the cores motivates the three-stage scenario for cluster formation illustrated in the top panel.}%
\end{figure*}

\section{Contribution to the Cosmic SFR Density}

Having quantified the large volumes of protoclusters, here we show that the SFR of all protoclusters combined accounts for a substantial $50\%$ of the CSFRD at $z=10$. In Fig.~\ref{fig:csfr} (upper panel) we plot the CSFRD for all galaxies (black curves) compared to the SFR density (SFRD) occurring exclusively in protoclusters (blue curves) and cores (red curves). The ratios of these curves (lower panel) show the fractional contributions of protoclusters and their cores to the total CSFRD. The closely related, absolute and fractional stellar mass densities are shown in the inset of each panel.

The global CSFRD predicted in the models are in general agreement with that derived observationally \citep{2005ApJ...619L..47S, 2012A&A...539A..31C, 2013ApJ...773...75O, 2015ApJ...810...71F}, with a broad peak at $z=1$--$3$ and decays on both sides. A slight offset in the normalization is present at high redshift, which is likely related to a known discrepancy between observational calibrations of galaxy SFR and stellar mass \citep[the latter is what is used in constraining the SAMs; see Fig.~11 in][]{2014ARA&A..52..415M}. For most of what follows, this will not affect the picture since the normalization would be canceled out.

In the models, the peak height of the SFRD for protoclusters is a factor of $5$ lower than that for the Universe as a whole. The peak of the SFRD for cores is another factor of $5$ lower, reinforcing the fact that for a full census of cluster formation it is insufficient to study only the most massive progenitor halos.

To understand the shapes of the SFRD curves, it is informative to consider the growth of halos in different environments in terms of their mass accretion rate \citep[MAR;][]{2015MNRAS.446..521S} and star-formation efficiency ($\rm SFE \equiv SFR/MAR$) as function of halo mass. The MAR increases monotonically with halo mass and density of the surrounding environment as the gravitational potential well is deepened. The SFE, however, peaks in halos of about $10^{12}\ \rm M_{\odot}$ due to various feedback processes at both lower and higher masses \citep[][]{2013ApJ...762L..31B,2013MNRAS.428.3121M}. This explains the rise and fall seen for all three SFRD curves. In more overdense regions, the characteristic halo mass grows more rapidly, reaching the peak SFE at $10^{12}\ \rm M_{\odot}$ earlier ($z\sim5$ for the cores \citep{2013ApJ...779..127C} compared to $z\sim1$ for the ensemble Universe \citep{1974ApJ...187..425P}). This leads to a similar, albeit less dramatic trend seen in the peak redshifts in each of the SFRD curves ($z_{\rm peak}\sim3$, 2.5, 2 for cores, protoclusters, and the global SFRD, respectively). The late-time decline in the SFRD is more rapid for both protoclusters and cores, reflecting the early quenching of galaxies in massive halos and dense environments.

Although (proto)clusters are made up of only $6\%$ of dark matter and baryons in the Universe (Fig.~\ref{fig:r_l}), at high redshift, they could be dominating the CSFRD. This is shown with the blue curves in the lower panel of Fig.~\ref{fig:csfr}. The fractional CSFRD in (proto)clusters is about $1\%$ at $z=0$ and increases to $20\%$ at $z=2$ and $50\%$ at $z=10$. Distant protoclusters may thus be important in driving the early history of cosmic star-formation.

Based on the resulting cosmic stellar mass evolution in these different environments as shown in the insets in Fig.~\ref{fig:csfr}, we find that galaxy clusters formed $50\%$ of their total stellar mass by $z=2$, about 2~Gyr before the Universe as a whole. If we consider the whole 4~Gyr ``Cosmic Noon'' epoch at $1<z<4$, the Universe and present-day clusters formed about $50\%$ and $75\%$ of their total stellar mass, respectively.

Although the CSFRD predictions in the G13 and H15 SAMs are very similar, interesting differences are seen when comparing the stellar growth in the cores. In H15 the fractional CSFRD in cores declines over time for the entire redshift range, while in G13 it rises after $z=2$. This is caused by the different implementations of the environmental effects operating in group- and cluster-size halos. This comparison shows that a detailed study of particularly the star formation history of protocluster cores could provide important constraints for modeling environmental processes.

\section{Internal Evolution of Galaxy Clusters}
We now investigate the internal evolution of the star formation and mass assembly of clusters. Fig.~\ref{fig:internal_evolution} shows the total SFR per protocluster as function of redshift averaged over our sample in the middle panel, and the fractional SFR (black curves) and $M_{\star}$ (red curves) in the core halo in the lower panel. We identify three distinct epochs during the history of cluster formation (as illustrated in the top panel):

\begin{enumerate}
  
\item 
From $z\gtrsim 10$ to $z\sim5$, galaxy growth in protoclusters appears to begin in an ``inside-out'' manner. This is based on the relatively large fractions of the SFR and $M_{\star}$ found in cores compared to protoclusters as a whole (the qulatative picture would stay the same if we define the core as a slightly extended region instead of a single halo). Although the cores represent only a tiny fraction of the protocluster volume at these redshifts, initially they dominated the SFR due to the higher MAR of these massive halos. On average, these halos reached peak SFE at $z\sim5$ when their masses approached $10^{12}\ \rm M_{\odot}$  \citep{2013ApJ...779..127C}. By this time, the SFR gradient across the entire protocluster flattens as more halos grow to near peak SFE, resulting in a drop in the fraction of the SFR and $M_{\star}$ associated with the cores to about $20\%$. The core remains dominant at a later cosmic time if seen in galaxy tracers of a higher limiting mass (dashed lines). This implies a relatively top-heavy stellar mass function in dense regions.
 
\item 
Between $z\sim5$ and $z\sim1.5$, the entire Lagrangian volumes of protoclusters contain numerous halos of $10^{11}$--$10^{12.5}\ \rm M_{\odot}$. This allows protocluster galaxies to grow at a total SFR of about $1000\ \rm M_{\odot}\ yr^{-1}$ for a prolonged period of time, which contributes to about $65\%$ of the total stellar mass seen in present-day clusters. Depending on the magnitude of the initial overdensity, some protoclusters may already contain significant group- or cluster-sized cores near the end of this epoch. These cores would be the first regions to show evidence of galaxy quenching or dense intracluster gas.
  
\item 
After $z\sim1.5$, the fraction of  $M_{\star}$ contained in protocluster cores starts to increase as the cluster is being assembled. The fraction of SFR in the cores lags behind its mass growth, implying that the growth of the cores is achieved mainly by incorporating externally-formed stars from in-falling galaxies. The violent gravitational collapse (Fig.~\ref{fig:density_profile} and~\ref{fig:r_l}) proceeds in an inside-out manner as the inner shells of a centrally peaked overdensity turn around before the outer shells. In this epoch galaxy quenching is enhanced through multiple channels, including gravitational heating, AGN feedback, group pre-processing, and various types of satellite quenching processes like starvation, ram-pressure stripping, and tidal disruption.
\end{enumerate}

Due to the extreme hierarchical nature of cluster assembly, the far majority of the stars in present-day clusters formed in the extended protocluster regions, mainly during the second phase outlined above. Only $15\%$ of the stellar mass formed ``in-situ'' in cores (this calculation takes into account mass loss during stellar evolution), which makes the core halos underrepresentative during the main epoch of cluster (galaxy) growth at Cosmic Noon.

\section{Discussion}

Based on two recent SAMs, we have demonstrated in this Letter that the fraction of the cosmic SFR density associated with the formation of present-day clusters is as high as $20\%$ at $z=2$ and $50\%$ at $z=10$. Protocluster galaxies are thus a nearly dominant population at Cosmic Dawn, and remain significant at Cosmic Noon.

We outlined three stages that describe the early history of cluster formation, which began with an inside-out growth phase from $z\gtrsim10$ to $z\sim5$, followed by an extended star-formation phase at $z\sim5$--$1.5$, and a violent infalling and quenching phase at $z\sim1.5$--$0$.

These phases are in qualitative agreement with tentative observational evidence. For example, \cite{2016ApJ...822....5I} reported a compact, $60$ physical kpc diameter overdensity of 8 dropout galaxies at $z\sim8$, presumably a prototypical, massive protocluster core in the first phase \citep[see also][]{2012ApJ...746...55T}. At $z\sim2$--$6$, a few dozen structures have been found that resemble the general properties of protoclusters expected during the second phase \citep[see a review in][]{2016A&ARv..24...14O}. Besides large, $10$--$20$ Mpc-scale overdensities of star-forming galaxies, some of them already contain massive cores. An example of the latter is the X-ray cluster at $z=2.5$ \citep{2016ApJ...828...56W} found at the center of a much more extended overdensity of star-forming galaxies \citep{2014ApJ...782L...3C}. Last, numerous clusters found at $z=0$--$1.5$ exhibit signatures of quenching and ongoing merging activities in the third phase of cluster formation \citep[][]{2005ApJ...624L..73G,2006ApJ...648L.109C}.

Our CSFRD predictions are consistent with a wide range of earlier studies. Using the abundance matching technique, \cite{2013ApJ...770...57B} showed that the SFH of cluster-sized halos peaked earlier, and reached higher values of SFR compared to lower mass halos. Stellar population synthesis applied to observational data of cluster galaxies implies high formation redshifts of their stellar content \citep[e.g.,][]{1999ApJ...518..576P,2009ApJ...690...42M,2010ApJ...709..512R}. Direct observations of $z\gtrsim2$ protoclusters have also shown a total SFR of hundreds to a few thousands $\rm M_{\odot}\ yr^{-1}$ in these extended regions \citep[][]{2005ApJ...626...44S,2014ApJ...789...18K,2015ApJ...808L..33C}. The large sizes and shapes of the protoclusters shown in Fig.~\ref{fig:popcorn} are also in qualitative agreement with the distribution of galaxies and gas absorption seen toward protoclusters \citep[][]{2016ApJ...817..160L,2016ApJ...833..135C,2017ApJ...835..281M}.

Our findings are natural consequences of the fact that cosmic structures and their constituent galaxies emerged first in highly biased regions. Based on their correlation length, \citet{2004ApJ...611..685O} concluded that the descendants of bright dropout galaxies at $z=4$--$5$ should be found in clusters today. At $z=9$--$10$, the number density of bright dropout galaxies \citep{2016ApJ...830...67B} is similar to that of present-day clusters.

The fact that the relative contribution from protoclusters to the CSFRD increases with redshift also implies that these structures played a significant role in cosmic reionization. Scaled from our SFR density result, nearly half of the hydrogen ionizing photons produced by massive stars in galaxies at $z>6$ originate from within the $5\%$ of cosmic volume occupied by protoclusters. \cite{2005astro.ph..7014C} suggested, based on the galaxy luminosity function of local clusters, that the cluster dwarf galaxy population might have been capable of reionizing the Universe on its own. Furthermore, if quasar activity at these redshifts closely traces overdense regions, the fractional ionizing radiation budget in protoclusters would be even higher. The popcorn-like structures shown in Fig.~\ref{fig:popcorn} are reminiscent of the ionizing bubbles seen in overdense regions in simulations of cosmic reionization \citep[][]{2003MNRAS.343.1101C,2007MNRAS.377.1043M,2014MNRAS.439..725I}. For these reasons, protoclusters may have been pivotal in driving the timing and topology of cosmic reionization.

In conclusion, our results show that galaxy protoclusters, while being rare in number density, are likely the leading sites of galaxy formation in ending the Dark Ages and driving the cosmic star-formation history in the first 2~Gyr. Deep galaxy surveys that sample a representative volume of the high-redshift Universe (e.g., CANDELS: \cite{2011ApJS..197...35G}, CLASH: \cite{2017arXiv170502265M}, and the Frontier Fields: \cite{2017ApJ...837...97L}) are probing relatively biased objects with massive $z=0$ descendants. Upcoming campaigns such as SuMIRe \citep{2014PASJ...66R...1T,2017arXiv170405858A}, HETDEX \citep{2008ASPC..399..115H,2015ApJ...808...37C}, and the WFIRST mission \citep{2013arXiv1305.5422S} will be able to survey the major stellar growth and test our SFR density prediction in extended protoclusters; deep observations with $JWST$ will be able to probe primordial protocluster cores and their connection to reionization.

\newcommand{\jcap}{JCAP}
\newcommand{\nar}{New Astron. Rev.}

\begin{acknowledgements}
YKC acknowledges support from NSF grant AST1313302 and NASA grant NNX16AF64G. RO received support from CNPq (400738/2014-7) and FAPERJ (E-26/202.876/2015). The Millennium Simulation databases used were constructed as part of the activities of the German Astrophysical Virtual Observatory (GAVO).
\end{acknowledgements}

\end{document}